\documentclass{aa}  
\usepackage{graphicx}
\usepackage{txfonts}
\usepackage{natbib}
\bibpunct{(}{)}{;}{a}{}{,} 
\begin{document}

   \title{Multispectral analysis of solar EUV images: \\
   linking temperature to morphology}

   \titlerunning{Multispectral analysis of solar EUV images}

   \author{Thierry Dudok de Wit
          \inst{1}
          \and
          Fr\'ed\'eric Auch\`ere\inst{2}
          }

   \offprints{T. Dudok de Wit}

   \institute{Laboratoire de Physique et Chimie de
    l'Environnement, UMR 6115 CNRS and University of Orl\'eans, 3A avenue de  
    la Recherche Scientifique, F-45071 Orl\'eans cedex 2, France\\
              \email{ddwit@cnrs-orleans.fr}
         \and
    Institut d'Astrophysique Spatiale, UMR 8617 CNRS and 
    University of Paris-Sud, B\^atiment 121, F-91405 Orsay, France\\
             \email{frederic.auchere@ias.u-psud.fr}
             }

   \date{Received 20 November 2006 / Accepted }

   \abstract
   {Solar images taken simultaneously at different wavelengths in the EUV are widely used for understanding structures such as flares, coronal holes, loops, etc. The line-of-sight integration and the finite spectral resolution of EUV telescopes, however, hinders interpretation of these individual images in terms of temperature bands. Traditional approaches involve simple visualisation or explicit modelling. We take a more empirical approach, using statistical methods.}
   {The morphology of solar structures changes with the wavelength of observation and, therefore, with temperature. We explore the possibility of separating the different solar structures from a linear combination of images.}
   {Using a blind source separation approach, we build a new set of statistically independent ``source'' images from the original EUV images. Two techniques are compared: the singular value decomposition and independent component analysis.}
   {The source images show more contrast than the original ones, thereby easing the characterisation of morphological structures. A comparison with the differential emission measure shows that each source image also isolates structures with specific emission temperatures.}
   {}  

   \keywords{solar EUV emission -- blind source separation}

   \maketitle
%
%

\section{Introduction}

\begin{figure}
\centering
    \includegraphics[width=0.99\columnwidth]{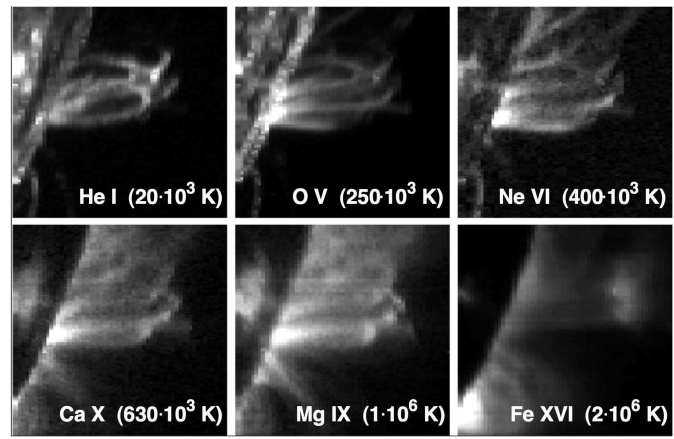} 
    \caption{The changing morphology of coronal loops depending on the temperature of observation. Images taken simultaneously at different wavelengths by the CDS spectrometer onboard SOHO near the solar limb on March 23, 1998 at 18:43:55 UT.}
    \label{fig_ica_cds}
\end{figure}

Since the early EUV images of the Sun, it has been known that the morphology of the transition region and the corona varies with the temperature of formation of the observed spectral lines. Typically, structures in the corona seem to be more diffuse when observed in hot lines and sharper in cool lines \citep{cheng80,gallagher98,feldman99}. This is particularly obvious when considering coronal loops, see Fig.~\ref{fig_ica_cds}.


The Extreme-ultraviolet Imaging Telescope (EIT) \citep{delaboudiniere95} onboard the SOHO spacecraft routinely takes images of the Sun in four passbands that are centred on intense lines of the EUV spectrum. These lines and their characteristics are summarised in Table~\ref{eit_properties}. The passbands of EIT are typically 1 nm wide, making multiple lines contribute to the total signal. As a consequence, the temperature response of EIT is a linear combination of the responses of the contributing lines, weighted by the spectral efficiency of the instrument. This mixing considerably complicates the interpretation of EIT images in terms of temperatures.

\begin{table*}[!htb]
\caption{EIT bandpasses}             
\label{eit_properties}      
\centering          
    \begin{tabular}{|c|c|c|l|}
       \hline
       wavelength & main line & peak temperature & appropriate for studying  \\
       \hline
       17.1 nm \ & Fe IX, Fe X & 1.3 MK & corona/transition region  \\
       19.5 nm \ & Fe XII & 1.6 MK & quiet corona  \\
       28.4 nm \ & Fe XV & 2.0 MK & active regions  \\
       30.4 nm \ & He II & 0.08 MK & chromospheric network \& coronal 
       holes  \\
       \hline
   \end{tabular}
\end{table*}

\begin{figure*}
\centering
    \includegraphics[width=0.75\textwidth]{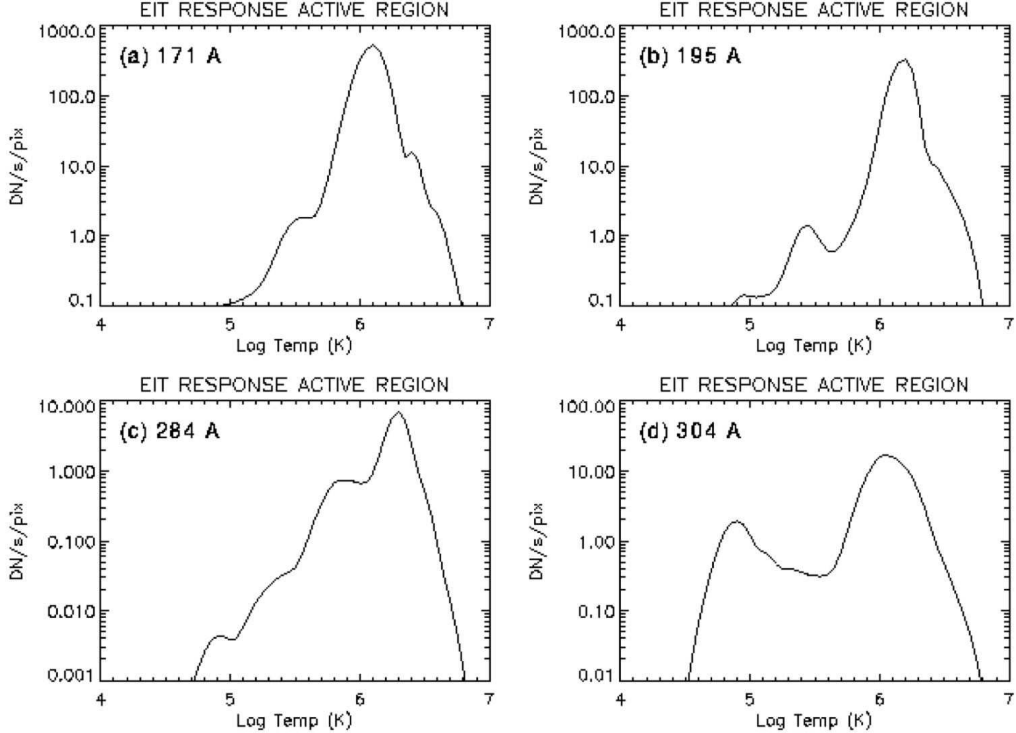} 
    \caption{Response of the EIT instrument to the plasma temperature, for an active region computed with an electron density of $10^9$ cm$^{-3}$ and using the differential emission measure from the CHIANTI model \citep{delaboudiniere95}.}
    \label{eit_response}
\end{figure*}

Figure~\ref{eit_response} shows the EIT temperature response for an active region. The 17.1, 19.5, and 28.4 nm passbands are strongly peaked around 1 MK, 1.5 MK, and 2 MK, respectively, but they also have secondary maxima and low-level extensions. The high-temperature wing of the 19.5 nm passband is due to an Fe XXIV line at 19.2 nm. This line, which is generated at 15.8 MK, is responsible for the very high sensitivity of the EIT 19.5 nm passband to flare plasmas. The low-temperature wing of the 28.4 nm passband is due to a group of transition region lines of Si VII and VIII. The 30.4 nm passband exhibits two peaks due to the contribution of two very different lines: the 30.38 nm line of He II and the 30.32 nm line of Si XI. The latter is generated in the corona at 1.1 MK and is responsible for the high-temperature peak, while the former is generated around 80 kK in the lower transition region.

The usual approach for deconvolving the temperature response is based on a generalisation of the differential emission measure (DEM) technique that was originally developed for narrow band spectra. This is done by replacing the contribution functions of individual spectral lines by the effective contribution functions of the passbands. Similar DEM inversions have already been successfully achieved with EIT images by \citet{cook99}, even though the lack of spectral coverage makes the problem underdetermined. 

Here, we consider the problem from a rather unconventional and more empirical point of view. Since each EIT image results from a mixture of various lines at different temperatures, we look for a linear combinations of images (called ``source images'') that are less redundant than the original ones and, in this sense, more contrasted. In doing so, we hope to disentangle the contributions of some of the spectral lines. This is a blind source separation (BSS) problem \citep{hyvarinen00}, which has generated considerable attention in astronomical image analysis and in speech processing. Here, as in most BSS problems, the source images are estimated using only the statistical properties of the EIT images. Such a statistical approach may seem quite empirical, yet it has several advantages:
\begin{itemize}
    \item in stereoscopic image analysis, the identification of structures, such as filaments and bright points, requires highly contrasted images. Multispectral analysis will be useful for preprocessing images from the dual STEREO probes. 
    
    \item for predicting space weather events, it is often more
    important to have fast access to approximate and empirical
    quantities (or proxies) than to computationally more costly physical quantities.
    
    \item  future missions, such as the Solar Dynamics Observatory, will produce unprecedentedly large amounts of multispectral images that will require fast automated analysis for data mining. 
    
    \item many multispectral analysis techniques often allow denoising
    as a by-product.
\end{itemize}

We consider two well known BSS techniques: the singular value decomposition (SVD), which is closely related to principal component analysis, and independent component analysis (ICA).  Although the source images are extracted by using statistical criteria only, they capture solar structures that differ in their morphology and are emitted in specific temperature bands. We interpret them visually, but also quantify them by comparing them to DEM maps. 

The method and the necessary preprocessing of the images are described in Sect.~2 and 3. The methods are first illustrated in Sect.~4 with synthetic images and then applied to solar images in Sect.~5. The results are discussed in Sect.~6, followed by a comparison with DEM maps. Finally, an application to multispectral imaging is presented in Sect.~7, followed by conclusions.


\section{Methodology}

In its usual operation mode, EIT routinely takes one image in each its four bandpasses several times per day.  These four images are generally taken a few minutes apart, thus enabling a multispectral study of structures that evolve on time scales longer than 30 minutes. The complex physical interpretation of these spectral lines has been discussed in detail by \citet{delaboudiniere95}, \citet{moses97}, \citet{aschwanden04}, and many others.  Each image is $1024 \times 1024$ pixels large, corresponding to a resolution of 2.6 arcsec per pixel.

Our analysis is based on the premise that each EIT image receives contributions stemming from physical processes whose spectral signatures and spatial distribution differ.  The three working hypotheses are:
\begin{enumerate}
    \item  the contributions from the different processes are mixed linearly;
    \item  their mixture is instantaneous (or non convolutive);
    \item  these sources are spatially independent, i.e. they have different spatial distributions.
\end{enumerate}
The first hypothesis is not perfectly satisfied insofar as the EUV spectrum is a blend of optically thin and thick lines. Only the He II line, however, is optically thick. The radiation transfer of that line in the corona is weak, so that linearity is still a reasonable working hypothesis. The second hypothesis amounts to saying that the Sun did not significantly evolve between the four snapshots.  The present cadence of EIT excludes the multispectral observation of fast transients such as eruptions, but future instruments will get much closer to this kind of an instantaneous picture. The last hypothesis is more difficult to evaluate. Solar structures with different temperatures tend to have different spatial distributions when they are observed at the limb (see for example Fig.~\ref{fig_ica_cds}), but nothing prevents the same line of sight from covering a broad range of temperatures, especially when it crosses the disk. The performance of BSS methods in the presence of non-independent sources  is a difficult problem, which can be alleviated by using more elaborate techniques such as multidimensional ICA \citep{hyvarinen00}. There is empirical evidence, however, for the separation of the sources to remain valid even if the condition of independence is not fully satisfied, provided that a significant fraction of the sample (of the pixels, in our case) captures structures that are independent enough to provide leverage to the method. Blind source separation methods are acceptable for the four-wavelength images of EIT, but will be more relevant for future instruments with more spectral channels.

Let $\left\{ \mathbf{y}_n \right\}_{n=1,2,3,4} $ denote four row
vectors containing the image intensities at each wavelength; images are 
thus stored in arrays. We assume that each observable is a linear combination of four \textit{source terms} $\left\{ \mathbf{s}_n
\right\}_{n=1,2,3,4} $
\begin{equation}
    \label{decomp_ica}
\mathbf{y}_n = \sum_{m=1}^4 A_n^m \mathbf{s}_m \ \ , \ \ \ n=1,2,3,4 \ ,
\end{equation}
where $\mathbf{A} = \left\{ A_n^m \right\}$ is the mixing matrix. Neither the mixing matrix nor the source terms are known a priori. The purpose of BSS is to determine both using the least possible information. This problem is strongly underdetermined, so the assumptions need to be clearly stated. For computational reasons, the number of source terms must be equal to or smaller than the number of observables. Solutions can still be found in the other case, but unicity is no longer guaranteed and the estimation becomes considerably more demanding. As we shall see, however, a small number of sources is partly justified by the strong redundancy of the four EIT images.

The most widely used technique for BSS is the
SVD \citep{golub96}, also known as the Karhunen-Lo\`eve transform, which is closely related to principal component analysis \citep{chatfield95}.  The SVD generates a new set of source images that are totally uncorrelated, i.e.,
\begin{equation}
\mathsf{E}(\mathbf{s}_n \mathbf{s}_m) = \left\{
\begin{array}{lll}
    0 & \textrm{ if } & n \neq m \\
    \sigma_n^2 & \textrm{ if } & n=m
\end{array}
\right. \ ,
\label{eq_variance}
\end{equation}
where $\mathsf{E}(\ldots)$ denotes expectation, and the energy $\sigma_n^2$ is the sum of squared intensities of image $n$.  The mixing matrix is unitary ($\mathbf{A}^T \mathbf{A} = \mathbf{A} \, \mathbf{A}^T = \mathbf{I}$, where $\mathbf{I}$ is the identity matrix) and can therefore be interpreted as a rotation matrix.  This decomposition is unique and the source terms are conventionally rank-ordered by decreasing magnitude of energy.  The sources with the highest energy are often of prime interest, since they capture salient features of the images.  The SVD can be understood as the most efficient way of removing redundancy from the observations, using a decomposition into four uncorrelated images. 

The ICA is a recent and powerful generalisation of SVD, which extracts source terms that are not only uncorrelated but also independent \citep{comon91, hyvarinen00, hyvarinen01}
$$
p(\mathbf{s}_m,\mathbf{s}_n) = p(\mathbf{s}_m) \; p(\mathbf{s}_n),
$$
where $p(\ldots)$ is the probability distribution function (PDF).  Independence is a stronger constraint than decorrelation, since it amounts to decorrelating all high-order statistical dependencies, in addition to the second-order one. 

For images that have a Gaussian PDF, the ICA is equivalent to the SVD. Gaussian images, however, tend to be the exception in astronomy; the departure from a Gaussian PDF is precisely exploited by the ICA to improve the separation of the sources. This will be illustrated shortly in Sect.~\ref{sec:synthetic}. It can be shown  that finding the independent components is equivalent to finding the set of images whose mutual information is minimised \citep{hyvarinen01}.  This property partly explains the recent popularity of ICA, which usually outperforms the SVD in BSS problems.

Numerous successful applications of ICA to blind deconvolution and to hyperspectral imaging have been reported. Several variants have also been proposed to tailor the method to physical constraints, see for example \citet{hyvarinen01}.  In astrophysics, \citet{nuzillard00} applied ICA to galactic image reconstruction, while \citet{baccigalupi00}, \citet{maino03}, and \citet{donzelli06} successfully used ICA to extract the cosmic microwave background from COBE data and \citet{lu06} to compress galaxy spectra. In solar physics, \citet{cadavid05} used ICA to investigate global modes in solar magnetic field fluctuations.


\section{Preprocessing}

The estimation of the SVD is computationally straightforward, since it amounts to diagonalising the $4 \times 4$ covariance matrix of the observations $\left\{ \mathbf{y}_n \right\}$.  The computation of the independent components requires an iterative scheme for which we use the FastICA algorithm \citep{hyvarinen00}.

Noise affects BSS, so it is important to know its properties beforehand. For sufficiently high intensities, the signal of the CCD camera tends to have a Poisson statistic for it is strongly affected by photon noise.  The variance can then be stabilised by applying the generalised Anscombe transform \citep{starck06}: if $y_m$ is the number of photons counted in pixel $m$, then the transform $y_m \longrightarrow \sqrt{y_m}$ provides a new set of pixel values that behave as if they were affected by stationary Gaussian noise.  We systematically apply this transform here.

Another problem is the lack of scaling invariance for the SVD and the ICA. The dynamic range of EIT images varies significantly with wavelength, so a normalisation is needed.  Our default procedure consists in centering (i.e. subtract from each image the average over all pixels) and reducing (i.e. normalise vs the standard deviation) each image.


\begin{figure*}
\centering
    \includegraphics[width=0.80\textwidth]{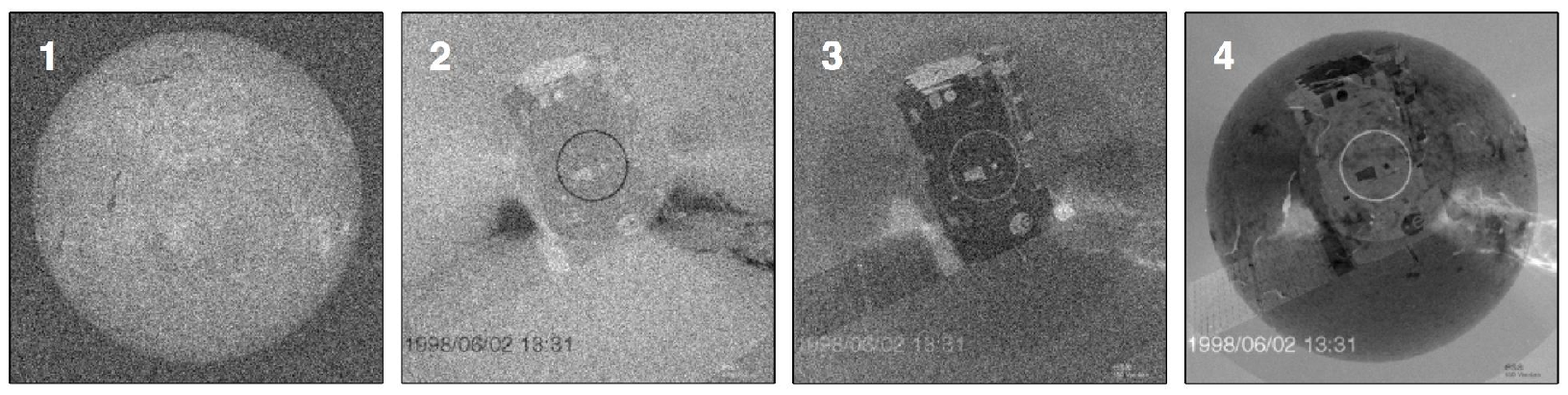}
    \includegraphics[width=0.80\textwidth]{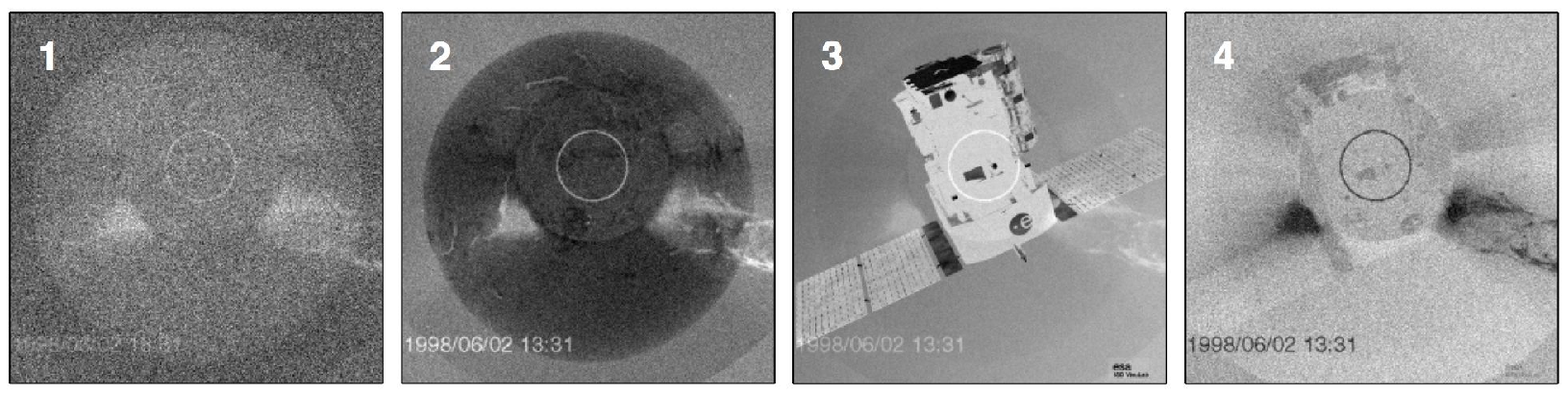} 
    \includegraphics[width=0.80\textwidth]{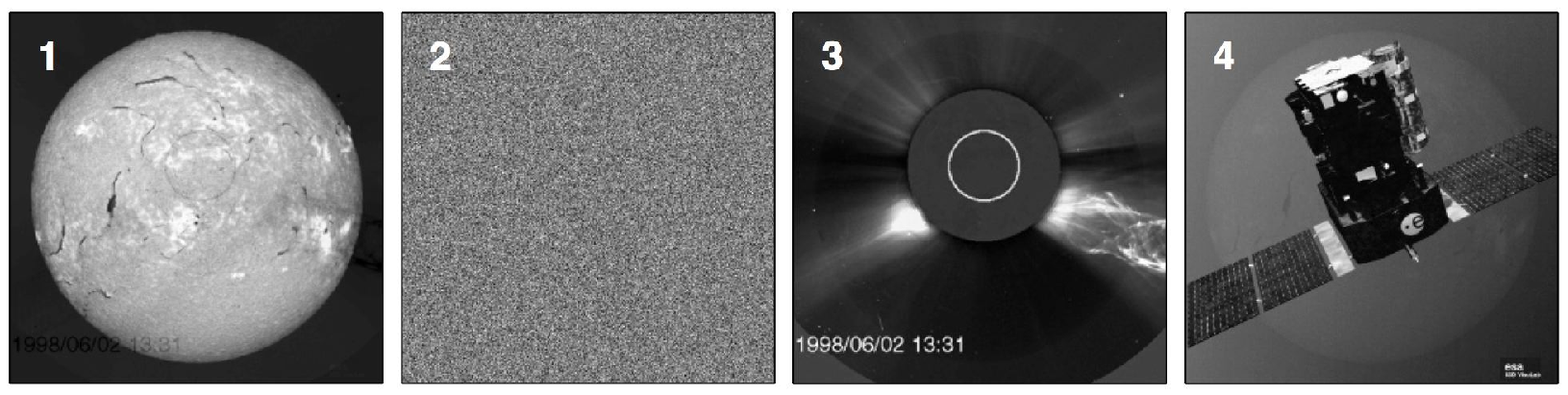} 
    \caption{From top to bottom: a random mixture of original
    images, their sources obtained by SVD, and by ICA. The original images are an image of the Sun taken in the H$\alpha$ line (ICA source nr. 1, courtesy of Big Bear Observatory), an image with white Gaussian noise (2), a coronagraph image of the Sun (3, from LASCO-C2), and a picture of the SOHO spacecraft (4, courtesy of ESA).}
    \label{fig_ica_mix}
\end{figure*}

\section{Example with synthetic images}
\label{sec:synthetic}

Let us first consider as a test case four $256 \times 256$ images, which we mix randomly before applying the SVD and the ICA. The noise properties are not known, so each original image is simply centred and reduced.  We subsequently build a set of four mixtures by combining the images linearly with random mixing coefficients.

The four mixtures are shown in the top row of Fig.~\ref{fig_ica_mix}. The application of the SVD to these data gives four source images that are displayed in the row below.  A weak visual improvement is apparent but none of the images looks pure yet.  The ICA (bottom row) in contrast recovers the four original images almost perfectly, owing to the additional leverage that is provided by the non-Gaussian PDF. These source images are unique up to a change in sign. 

A metric is needed here to rank the images without having to go through the subjective task of visualisation.  There is no such unique metric, but some helpful measures exist.  One natural measure for the SVD is the variance $\sigma_n^2$ of the images (see Eq.~\ref{eq_variance}), since both the images and the columns of the mixing matrix are orthogonal by construction.  

The information theoretical foundation of ICA suggests that the entropy might be an appropriate measure.  We consider the relative entropy or Kullback-Leibler distance $H_{KL}$ \citep{cover91}, which quantifies the amount of information we gain from the image with PDF $p(y)$ as compared to an image with PDF $q(y)$
\begin{equation}
H_{KL} = \int  p(y) \log_2 \frac{p(y)}{q(y)} \;  dy \ .
\label{eq:kullback}
\end{equation}
A natural choice for the reference PDF $q(y)$ is a Gaussian with the same mean and variance as the image of interest.  The relative entropy can then be interpreted as a measure of the deviation from Gaussianity; $H_{KL}=0$ if and only if $q(y)$ is a Gaussian.  

Table~\ref{table_entropy} below lists the entropies obtained from the images of Fig.~\ref{fig_ica_mix}. A discrete version of Eq.~\ref{eq:kullback} was applied, using histograms with 80 equispaced intervals. Note that the relative entropy of a mixture is generally lower than that of its sources, by virtue of the central limit theorem. The relative entropy of the source images obtained by SVD is relatively low.  The results are much more contrasted with ICA, which by definition searches for independent source images that extremalise the relative entropy.

\begin{table}[!htb]
	\caption{Relative entropy $H_{KL}$ associated with the images of Fig.~\ref{fig_ica_mix}. Units are bits.}
    \label{table_entropy}
    \centering
    
    \begin{tabular}{|l|cccc|}
	\hline
	image & 1 & 2 & 3 & 4  \\
	\hline
	mixture & 0.020 & 0.018 & 0.002 & 0.257  \\
	SVD sources & 0.002 & 0.151 & 0.092 & 0.078  \\
	ICA sources & 1.025 & 0.001 & 0.384 & 0.249  \\
	\hline
    \end{tabular}
\end{table}


\section{Example from solar images: results}

Let us now consider real data with four images taken by the EIT telescope on August 1st, 2003, at 19:00:14 (17.1 nm), 19:06:01 (28.4 nm), 19:13:41 (19.5 nm), and 19:20:05 (30.4 nm) UT. The Sun at that time exhibits some large but relatively stable active regions. The data are preprocessed in the usual way: the dark-current offset is subtracted, a flat-field correction is made, the Anscombe transform is applied, and finally the images are centred and reduced. 


\begin{figure*}
\centering
    \includegraphics[width=0.80\textwidth]{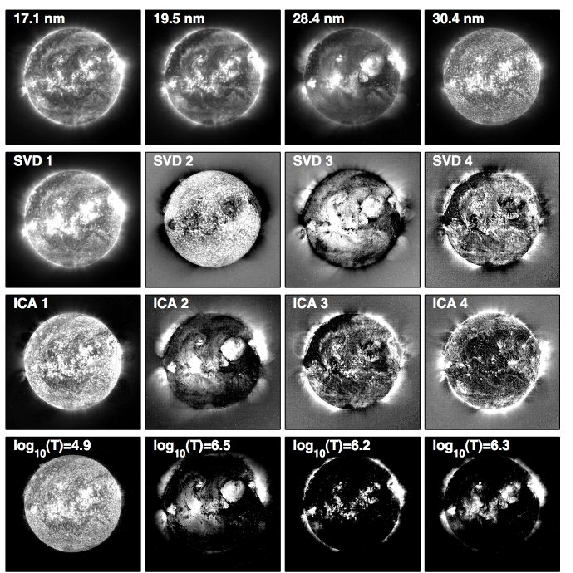}
    \caption{From top to bottom: original EIT images of the Sun
    (respectively at 17.1 nm, 19.5 nm, 28.4 nm, and 30.4 nm), the
    source images obtained by SVD and by ICA, and the DEM at four temperatures. 
    The intensity ranges
    from the 5\% percentile (black) to the 98 \% percentile (white).
    The data have been preprocessed as explained in the text.  Each
    image is 1024 $\times$ 1024 pixels in size. The SVD and the ICA images are 
    ranked by decreasing variance.}
    \label{fig_ica_all}
\end{figure*}


The pre-processed EIT images are displayed in the top row of Fig.~\ref{fig_ica_all}. The 17.1 to 28.4 nm spectral lines mostly probe the hot corona and active regions, whereas the 30.4 nm line reveals the much cooler network. The result of the SVD and the ICA are shown respectively in the second and the third rows of Fig.~\ref{fig_ica_all}.  Note that these images should not be interpreted as intensities, since pixel values can be negative. Also shown are DEM maps, determined for the characteristic temperatures of the EIT passbands given in Table~\ref{eit_properties}. The DEM was estimated from the same set of EIT images, for each pixel, using the procedure described in \citet{cook99}. 

The images are sorted here by decreasing variance.  The relative variance of the SVD sources is respectively 46~\%, 41~\%, 13~\%, and 0.05~\%.  The low value of the last suggests that most of the salient features are only captured by SVD images 1 to 3.  This result is a direct consequence of the redundancy of information in the four EIT images. The SVD can in this sense be considered as a lossy compression tool.

An inspection of the PDFs confirms that the ICA source images strongly depart from a Gaussian and yet are independent. This is summarised by the values of the relative entropy, see Table~\ref{table_entropy2}. The larger the entropy, the more non-Gaussian the distribution.

\begin{table}[!htb]
	\caption{Relative entropy $H_{KL}$ (in bits) of the images shown in Fig.~\ref{fig_ica_all}.}
    \label{table_entropy2}
    \centering
    \begin{tabular}{|l|cccc|}
	\hline
	image nr. & 1 & 2 & 3 & 4  \\
	\hline
	original images & 0.61 & 0.77 & 0.76 & 1.01  \\
	SVD sources & 0.42 & 0.13 & 0.13 & 0.12  \\
	ICA sources & 0.68 & 0.22 & 0.15 & 0.15  \\
	\hline
    \end{tabular}
\end{table}


\section{Example from solar images: interpretation}

The source images obtained by SVD and by ICA in Fig.~\ref{fig_ica_all} exhibit several interesting features that are easier to understand when we consider the mixing matrix. The rows of the inverse mixing matrix $\mathbf{A}^{-1}$ indeed tell us which linear combination of the EIT images is needed to build the source images.


\begin{figure*}
\centering
    \includegraphics[width=0.70\textwidth]{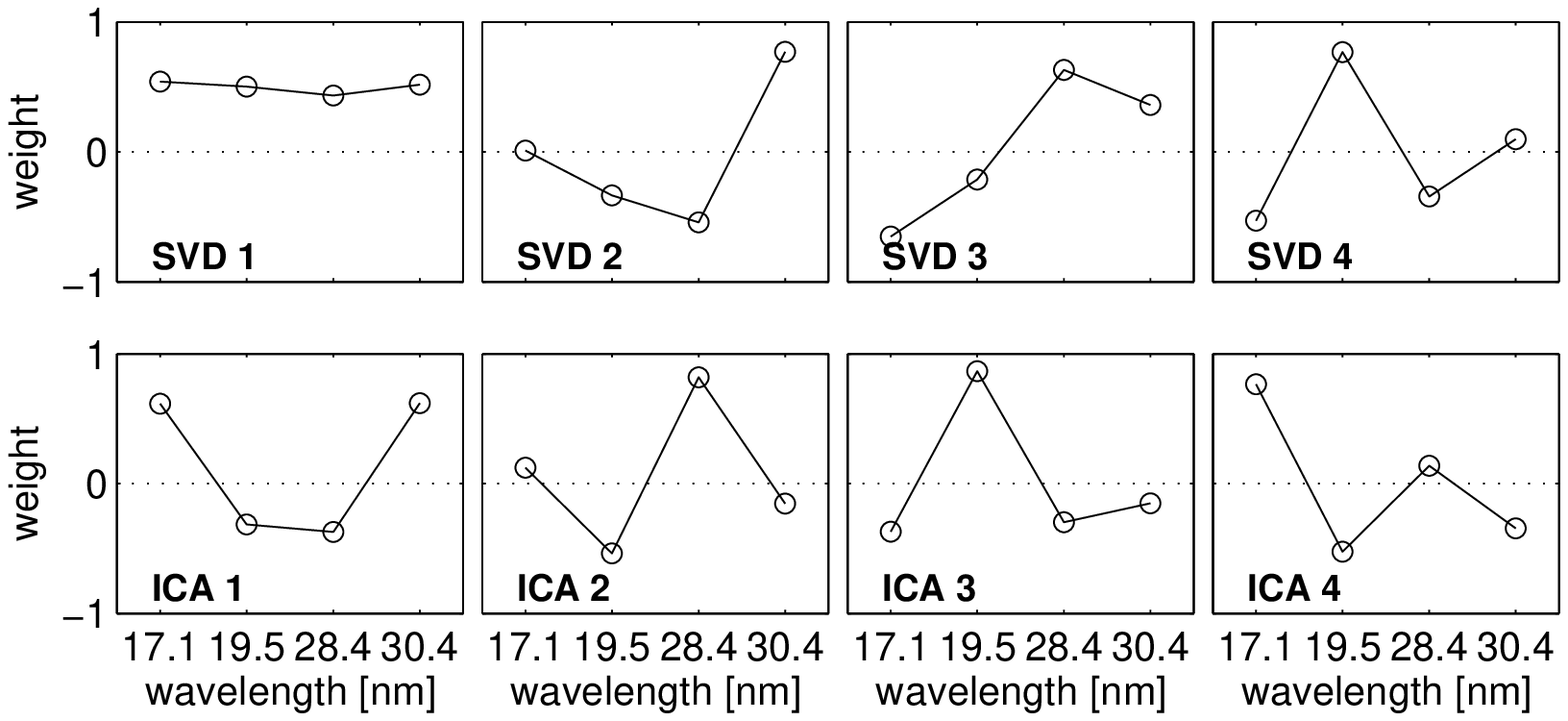} 
    \caption{The weighting coefficients applied to the original EIT
    images to obtain the source images by SVD (top row) and ICA (bottom
    row).  The numbering of the images corresponds to 
    Fig.~\ref{fig_ica_all}.}
\label{fig_ica_topo2}
\end{figure*}


The first SVD source image is obtained by averaging over all EIT channels, see Fig.~\ref{fig_ica_topo2}.  Since each channel is dominated by hot active regions, the last ones are enhanced by the averaging. The second SVD source image mostly captures the emission at 30.4 nm  minus that of the other lines. Not surprisingly, it emphasises the fine structure of the network. The fourth source, which only weakly contributes to the EIT images, mostly captures small-scale features including the shadow of the instrument mask and cosmic ray hits. 

The ICA source images differ from the SVD ones. The first source image, which is a combination of the images at 17.1 and 30.4 nm, minus the contribution at 19.5 and 28.4 nm, enhances the cold network, similar to the second SVD source image. A deeper inspection, however, reveals that the contribution from the hot coronal lines is subtracted by the ICA better than by the SVD. The second ICA source essentially captures the emission at 28.4 nm minus that at 19.5 nm (see Fig.~\ref{fig_ica_topo2}). This results in an enhancement of the active regions and an attenuation of the medium hot corona. Note that the source images are more contrasted than in the original EIT images. This is not surprising, since the ICA proceeds by maximising the information content of each image, while keeping all four images independent. 

A visual inspection of ICA source image 3 suggests that it captures the bright limb and structures seen in the medium hot corona. According to Fig.~\ref{fig_ica_topo2}, it indeed consists of the emission at 19.5 nm minus that of the two other hot lines. More quantitative evidence for this will be given in the next section. The fourth source image is the hardest to interpret, even though some of its features are reminiscent of what one would expect from the lower corona. As for the SVD, the contribution of that image is very small, so we omit it. We indeed checked that a reconstruction from three sources gives almost the same EIT images as before, while the reconstruction with two sources clearly remains incomplete.

The key result of the ICA (and to a lesser degree of the SVD) is the decomposition of the EIT data into source images that reveal structures with quite different morphologies. ICA source image 1 exhibits a fine structured network only on the disc, whereas source image 2 shows diffuse regions; image 3 shows fine structures both on the disc and in the corona. These images are qualitatively similar to the DEM maps (displayed in the bottom row of Fig.~\ref{fig_ica_all}), thereby suggesting the existence of a link between morphology and temperature.

The notion of morphology is vague, although widely used in the literature; see for example \citet{cheng80}. A wavelet analysis of solar EUV images \citep{delouille05} shows that the broadband power spectral density follows a power law: if $P(a)$ is the power spectral density at scale $a$, then $P(a) \propto a^{\beta}$. The occurrence of such a power law precludes the existence of a characteristic scale, except for the network in the cold lines. Our visual perception of morphology is closely associated with the value of the spectral index $\beta$. High values of $\beta$ imply steep power spectra and a predominance of large-scale structures. For random noise, $\beta$ equals zero. 

\begin{figure}
\centering
    \includegraphics[width=0.8\columnwidth]{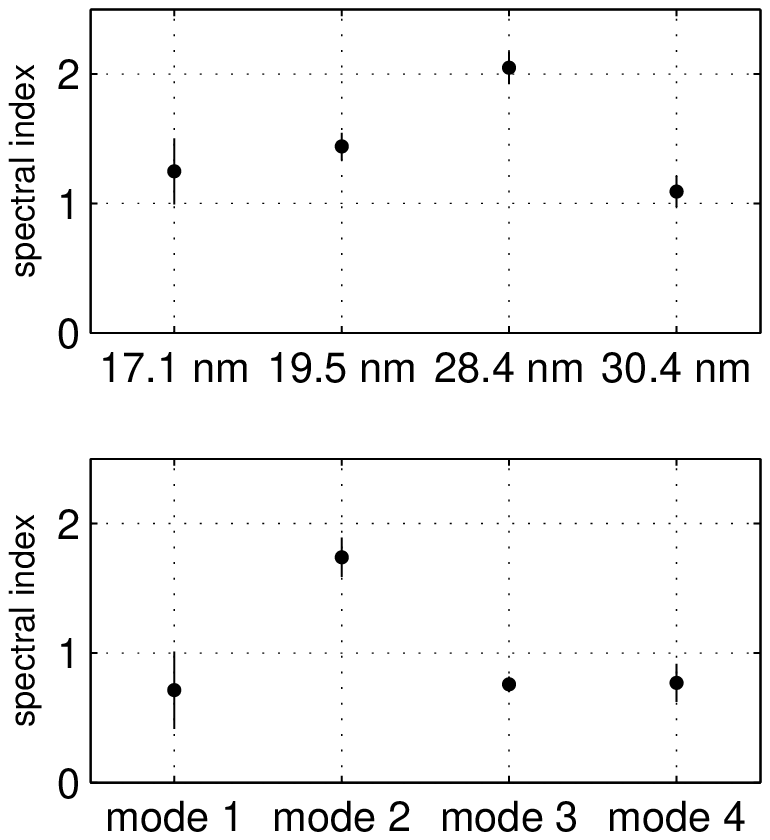} 
    \caption{Spectral indices $\beta$ estimated from the original EIT images (top) and from the four ICA modes (bottom). The indices are estimated by the wavelet transform, for scales ranging from 32 to 128 pixels. Error bars correspond to $\pm$ one standard deviation resulting from the power law fit.}
\label{fig_ica_index}
\end{figure}

Here we estimate the spectral index $\beta$ by computing the two-dimensional discrete wavelet transform over the solar disc, using third-order Daubechies wavelets \citep{mallat98}. The main advantage of this estimator is its resilience to trends. It has also been shown by \citet{abry95} that this estimator provides an unbiased and more robust estimate of the spectral index. The results are summarised in Fig.~\ref{fig_ica_index}.


The spectral indices of the original images confirm our visual perception of scale distribution. EIT images at 28.4 nm, which exhibit hot and diffuse structures in the corona, also have the highest spectral index. The lowest index arises at 30.4 nm, which indeed shows the network. The same ordering holds for the SVD (not shown) and the ICA source images. ICA mode 2, which captures diffuse structures, also has the highest index. Mode 1, which corresponds to the cold network, has the lowest index. Interestingly, the source images have lower spectral indices than the original ones, which corroborates the idea that source images are more informative (in the sense of being more structured) than their mixture.


\section{Comparison with the differential emission measure}

To put our qualitative results on firmer ground, we now compare the source images to DEM maps. The inversion procedure by \citet{cook99} was used to estimate the DEM pixel-wise, for a set of 20 temperatures ranging logarithmically from $10^{4.6}$ to $10^{6.5}$ K. 

To correlate DEM maps with the source images, one would traditionally use Pearson's correlation coefficient. We prefer Spearman's rank correlation coefficient \citep{press92}, which has the major advantage of being non-parametric and thus insensitive to any rescaling of the images by a nonlinear monotonic function. Indeed, Spearman's correlation coefficient compares the ranks of the image pixels and not their values. Both correlation coefficients have the same interpretation: 1 implies full correlation, -1 full anticorrelation, and 0 no correlation.

\begin{figure*}
   \centering
   \includegraphics[width=0.80\textwidth]{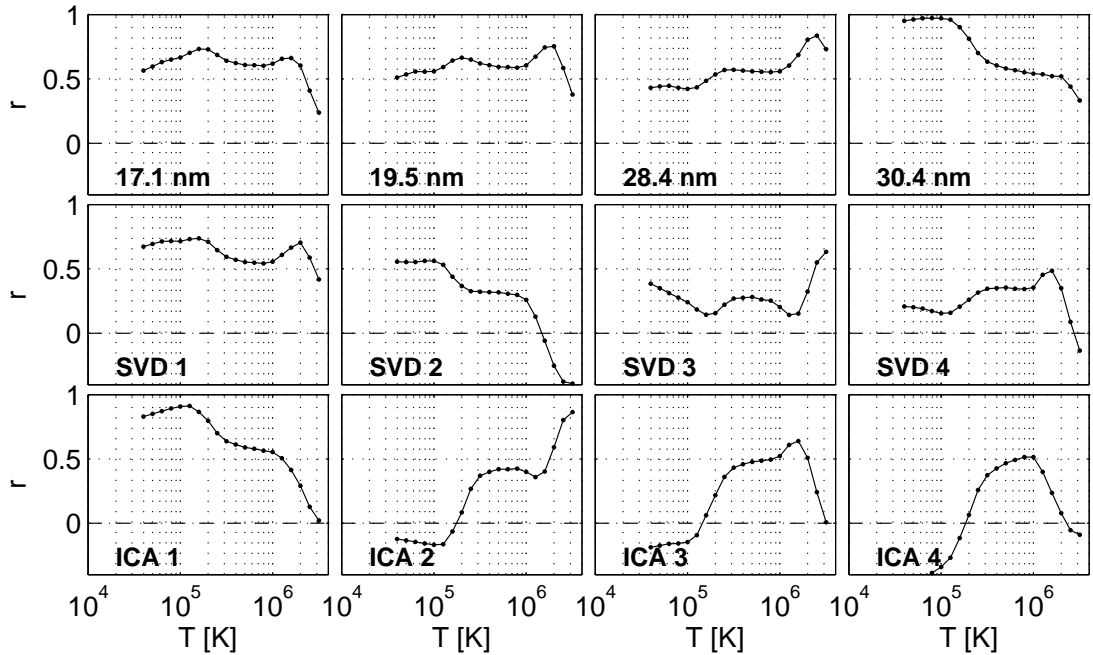}
   \caption{Spearman's rank correlation coefficient $r$ as a function of temperature $T$. The correlation is computed between the DEM and: original EIT images (top row), SVD source images (middle row), and ICA source images (bottom row). The ordering of the images is identical to Fig.~\ref{fig_ica_all}. }
\label{fig_ica_spearman}
\end{figure*}

Here, we estimate Spearman's correlation coefficient between each DEM map and a) the four EIT images, b) the four source images obtained by SVD, and c) the four source images obtained by ICA. The correlations are plotted in Fig.~\ref{fig_ica_spearman} versus the characteristic emission temperature. The mixture of lines that contribute to each EIT image is clearly attested by the flat and multimodal distribution of Spearman's rank correlation coefficient; see the top row of Fig.~\ref{fig_ica_spearman}. The highest peak generally coincides with the characteristic emission temperature of each spectral band. The smooth shape of the correlation is a consequence of the DEM reconstruction technique, which cannot resolve fine temperature bands. The transition region between $10^5$ and $10^6$ K should be disregarded, because the DEM maps in that range are merely extrapolated from the temperatures above and below. 

The SVD partly succeeds in separating solar features according to their characteristic temperature. Each image is now more strongly dominated by one particular temperature band, and yet source images 1 and 3 still exhibit double peaks. The key result is that a better separation is achieved by the ICA as compared to the SVD, since each source image reveals one single temperature band. For source image 1, the correlation is very similar to that of the original EIT image at 30.4 nm, with a large maximum corresponding to the cold lower transition region. In contrast to the EIT image, however, the high temperature contribution above $10^6$ K is now subtracted better. Source image 2 captures the hottest coronal regions better than any other. The resolving power of source image 3 is somewhat weaker, yet it remains the best image for extracting the medium hot corona. Source image 4 is more enigmatic, but since it peaks in the temperature band where the DEM is the least well defined, a physical interpretation is risky.

We conclude that the ICA succeeds in separating structures that have both specific morphologies and specific characteristic temperatures. It is remarkable that this separation is achieved only on the basis of statistical criteria, without invoking the complex physics behind solar spectroscopy. We believe that the reason for this successful separation comes from the intimate connection between the physical characteristics of solar structures observed in EUV and their morphological properties, which provide leverage to the ICA. The resulting source images are of course empirical. The Solar Dynamics Observatory, with its 9 wavelengths, will provide a different set of sources that are likely to provide an even better temperature separation. As already mentioned in the introduction, these empirical source images can be valuable input for space weather products, for stereoscopy, and for plain visualisation.


We must stress that all these results are reproducible in the sense that we obtain essentially the same combinations regardless of the period the data are taken from (quiet Sun versus active Sun).  The results are also robust with respect to the choice of the region, as long as the region covers a sufficiently complete ensemble of solar structures, including both the disc and the corona. This reproducibility is important since it means that the mixing matrix needs to be computed only once, regardless of the number of events considered.

Our results are obtained after taking the square root of the image intensity. Such a transform, however, does not conform with the linear relationship between source images. The connection between the ICA source images and different temperature bands of the DEM actually hardly changes if the Anscombe transform is not applied, or if we take the logarithm of the images (which is equivalent to assuming multiplicative sources). This resilience to the preprocessing of the images probably has to do with the occurrence, for each characteristic temperature, of structures at all intensities. As a consequence, the separation into different morphologies is rather insensitive to the scaling of the intensity. The clearest temperature separation, however, is obtained with the Anscombe transform.


\section{Multispectral imaging of the Sun}

\begin{figure*}
\centering
    \includegraphics[width=0.8\textwidth]{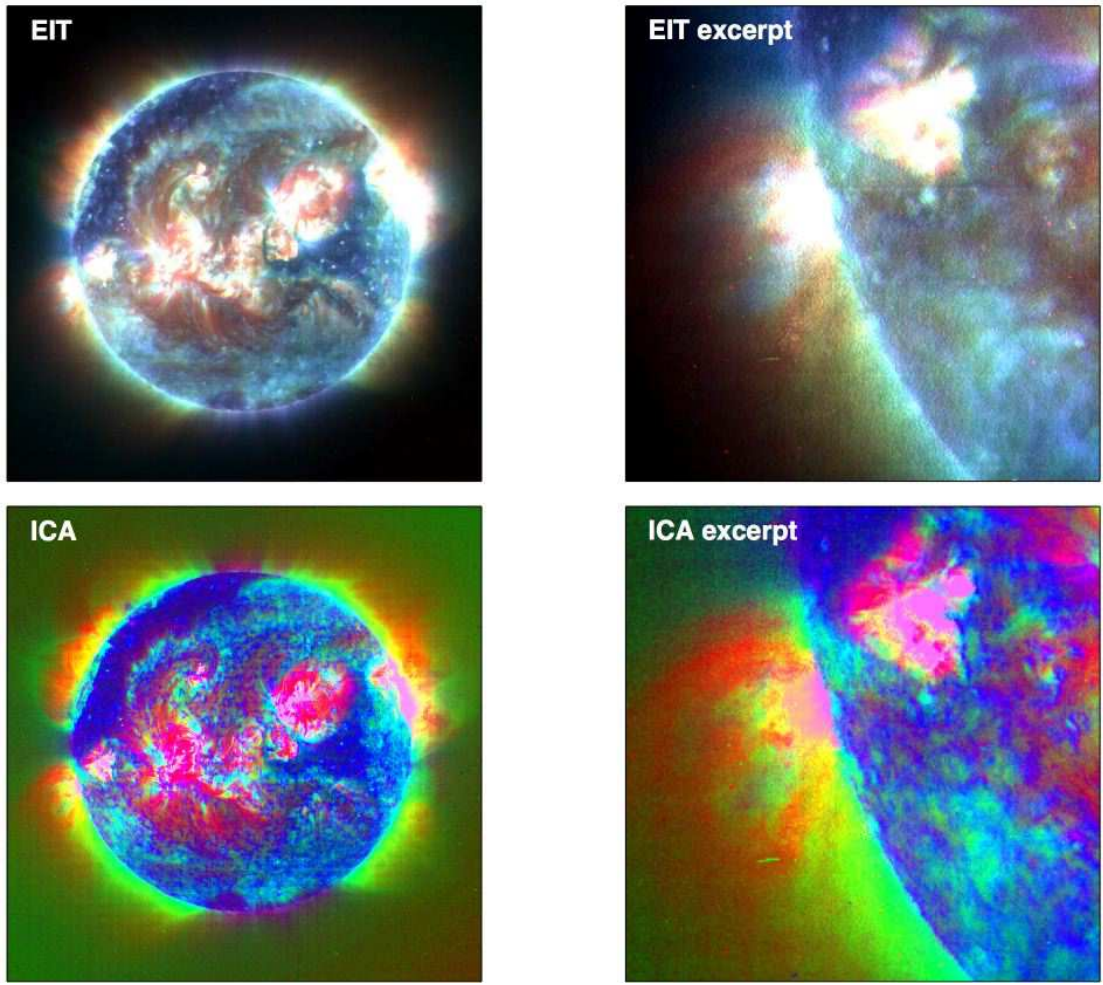} 
    \caption{Multispectral solar images obtained by assigning different image intensities to the red, green, and blue channels. The top images are obtained from the traditional combination of EIT images at 28.4 nm \, (red), 19.5 nm \, (green), and 17.1 nm \, (blue). The bottom images are obtained from ICA source images 2 (red), 3 (green), and 1 (blue). The right column shows an enlarged portion of the east limb.}
\label{fig_ica_multi}
\end{figure*}

An interesting by-product of the SVD and the ICA is a feature called multispectral imaging in the remote sensing of the Earth. The interpretation of multispectral images is often hampered by the difficulty visualising different wavelengths at once. A simple solution consists in combining three of the images into a single colour picture, by assigning different image intensities to the red, green, and blue channels. When applying this method to raw EIT images, using the red channel for 28.4 nm, the green one for 19.5 nm,  and the blue one 17.1 nm, we obtain the pictures shown in Fig.~\ref{fig_ica_multi}.


Most of the features in Fig.~\ref{fig_ica_multi} appear greyish because the information stored in the three colour channels is highly redundant. A less redundant combination of colours channels would be needed, for which the source images are an obvious solution. Figure~\ref{fig_ica_multi} shows the multispectral image obtained by combining ICA source images 1, 2 and 3; the ordering of the colours reflects our perception of the characteristic temperature of each source. The higher contrast eases the visual interpretation of different solar structures. Notice in particular how active regions now appear less saturated.  Although this method gives a purely qualitative view of the Sun, its widespread use in airborne surveying suggests that it can be a valuable tool as well for visualising multispectral solar images. The same technique can also be used for segmentation purposes and the robust detection of different regions such as coronal holes, see \citet{DdW06}.


\section{Conclusions}

The broadband spectral response of EUV imagers has stimulated the search for finding ways of extracting specific temperature bands from multispectral images. In this study, we used a BSS approach to tackle this problem in a statistical, and thus rather unconventional, way. Using independent component analysis, we show how multispectral images taken by EIT can be decomposed into a set of more contrasted source images that reveal solar structures with specific morphologies. A comparison with the DEM indeed confirms that each source image captures solar structures that have a specific emission temperature, thereby establishing a natural link between morphology and temperature. The most conspicuous regions that come out are the lower transition region, the million degree corona, and the hot corona. 

The source images are obtained on the basis of the statistical properties of the EIT images only and so cannot be interpreted as temperature maps, and yet their good agreement with the DEM suggests that they can be valuable for a visual inspection of EUV images. These source images are easy to compute, which makes this empirical approach well-suited to processing large volumes of multispectral data. The concept of BSS will become more relevant as future missions will be observing in more than four wavelengths simultaneously. 

One obvious improvement, which is in progress, consists in constraining the source images, as well as their mixing coefficients, to be positive. Another approach would consist in mixing ICA with a multiscale analysis by first decomposing the images into multiple scales using a wavelet transform and then performing the ICA.


\begin{acknowledgements}
      We thank the International Space Science Institute (ISSI, Bern), where this study was initiated, for its hospitality. The EIT and CDS teams are acknowledged for providing excellent data. SOHO is a joint project by ESA and NASA.
\end{acknowledgements}


\bibliographystyle{aa}
\bibliography{ica}

\end{document}